\documentclass[aps,prl,twocolumn,groupedaddress]{revtex4}
\usepackage{amsmath}
\usepackage{amssymb}
\usepackage{graphicx}
\usepackage{latexsym}
\usepackage{subfigure}
\begin{document}

%\preprint{}

\title{Helical Packings and Phase Transformations of Soft Spheres in Cylinders}


\author{M. A. Lohr$^1$, A. M. Alsayed$^{1,2}$, B. G. Chen$^1$, Z. Zhang$^{1,2}$, R. D. Kamien$^{1,3}$, A. G. Yodh$^1$}
%\thanks{}
\affiliation{$^1$Department of Physics and Astronomy, University of Pennsylvania, Philadelphia, PA 19104, USA}
\affiliation{$^2$Complex Assemblies of Soft Matter, CNRS-Rhodia-UPenn UMI 3254, Bristol, Pennsylvania 19007, USA }
\affiliation{$^3$School of Mathematics, Institute for Advanced Study, Princeton, NJ 08540, USA}

\date{\today}

\begin{abstract}
The phase behavior of helical packings of thermoresponsive microspheres inside glass capillaries is studied as a function of volume fraction.  Stable packings with long-range orientational order appear to evolve abruptly to disordered states as particle volume fraction is reduced, consistent with recent hard sphere simulations.  We quantify this transition using correlations and susceptibilities of the orientational order parameter $\psi_6$.  The emergence of coexisting metastable packings, as well as coexisting ordered and disordered states, is also observed.  These findings support the notion of phase transition-like behavior in quasi-1D systems.
\end{abstract}
%\keywords{}
\maketitle

The phenomenology of ordered phases and phase transformations in systems with low dimensionality is surprisingly rich.  While dense three-dimensional (3D) thermal packings can exhibit long-range order, for example, this trait is absent in purely one dimensional systems [1].   Complexities arise, however, when considering 3D systems confined primarily to one dimension (1D).   Investigation of order and phase behavior in quasi-1D thermal systems, therefore, holds potential to elucidate a variety of novel physical processes that have analogies with polymer folding [2], formation of supermolecular fibers in gels [3], and emergence of helical nanofilaments in achiral bent-core liquid crystals [4].

Packings of soft colloidal spheres in cylinders provide a useful model experimental system to quantitatively investigate order and phase transformations in quasi-1D.  At high densities, spheres in cylindrical confinement are predicted to form helical crystalline structures [5,6].  Evidence for such packings have been found in foams [7,8], biological microstructures [5], colloids in micro-channels [9] and fullerenes in nanotubes [10].  However, research on these systems has been limited to static snapshots and athermal media.  Recent simulations suggest that transitions between different helical ordered states [11,12] and between quasi-1D ordered and disordered states [13-15] should exist in thermal systems, but such transitions have not been investigated experimentally.  

In this Letter, we explore ordered and disordered structures in a quasi-1D thermal system of soft particles with adjustable volume fraction.  In particular, we create helical packings of thermoresponsive colloid particles in glass microcapillaries, we show theoretically that phases with long-range orientational order can exist in quasi-1D, we demonstrate experimentally that such phases with long-range orientational order exist at volume fractions below maximal packing, and we study volume-fraction induced melting of these orientationally ordered phases into liquid phases.  The orientational order parameters and susceptibilities that characterize these phases and this crossover are measured and analyzed.  Coexisting regions of ordered and disordered states and coexisting ordered domains with different pitch and chirality are observed at these crossover points.  Such coexistence effects suggest the presence of abrupt or discontinuous volume-fraction driven transitions in quasi-1D structures.  Interestingly, these orientationally ordered phases in quasi-1D share physical features with orientationally ordered phases observed [16,17] and predicted [18] in 2D.  

The experiments employed aqueous suspensions of rhodamine-labeled poly-N-isopropylacrylamide (NIPA) microgel spheres (polydispersity $\le3\%$) with diameters which decrease linearly and reversibly with increasing temperature [19].  The unique thermoresponsive characteristics of NIPA microgel particles provide a means to explore the phase behavior [17,20-21] of soft spheres in quasi-1D as a function of volume fraction.  Borosilicate glass tubes (McMaster Carr) were heated and stretched to form microcapillaries with inner diameters comparable to NIPA microsphere diameter.  An aqueous suspension of NIPA microspheres was then drawn into the capillaries; subsequently, the capillary ends were sealed with epoxy and attached to a glass microscope slide.  Samples were annealed at 28$^\circ$C to permit uneven packings to re-arrange at low volume fraction, thereby creating stable high-volume fraction packings when returned to lower temperatures.  

The samples were imaged under a 100$\times$ oil-immersion objective (N.A. = 1.4) using spinning-disk confocal microscopy (QLC-100, Visitech, International).  Resulting images depict 75$\mu$m long segments of densely packed regions which span at least several hundred microns.  An objective heater attached to the microscope (Bioptechs) permitted control of the sample temperature to within $\pm$0.1$^\circ$C.  Standard particle tracking routines [22] were employed to identify particle positions from three dimensional image stacks and 2D cross-sections.  

\begin{figure}
\centering
\includegraphics*[viewport=10 0 450 340, scale=0.53]{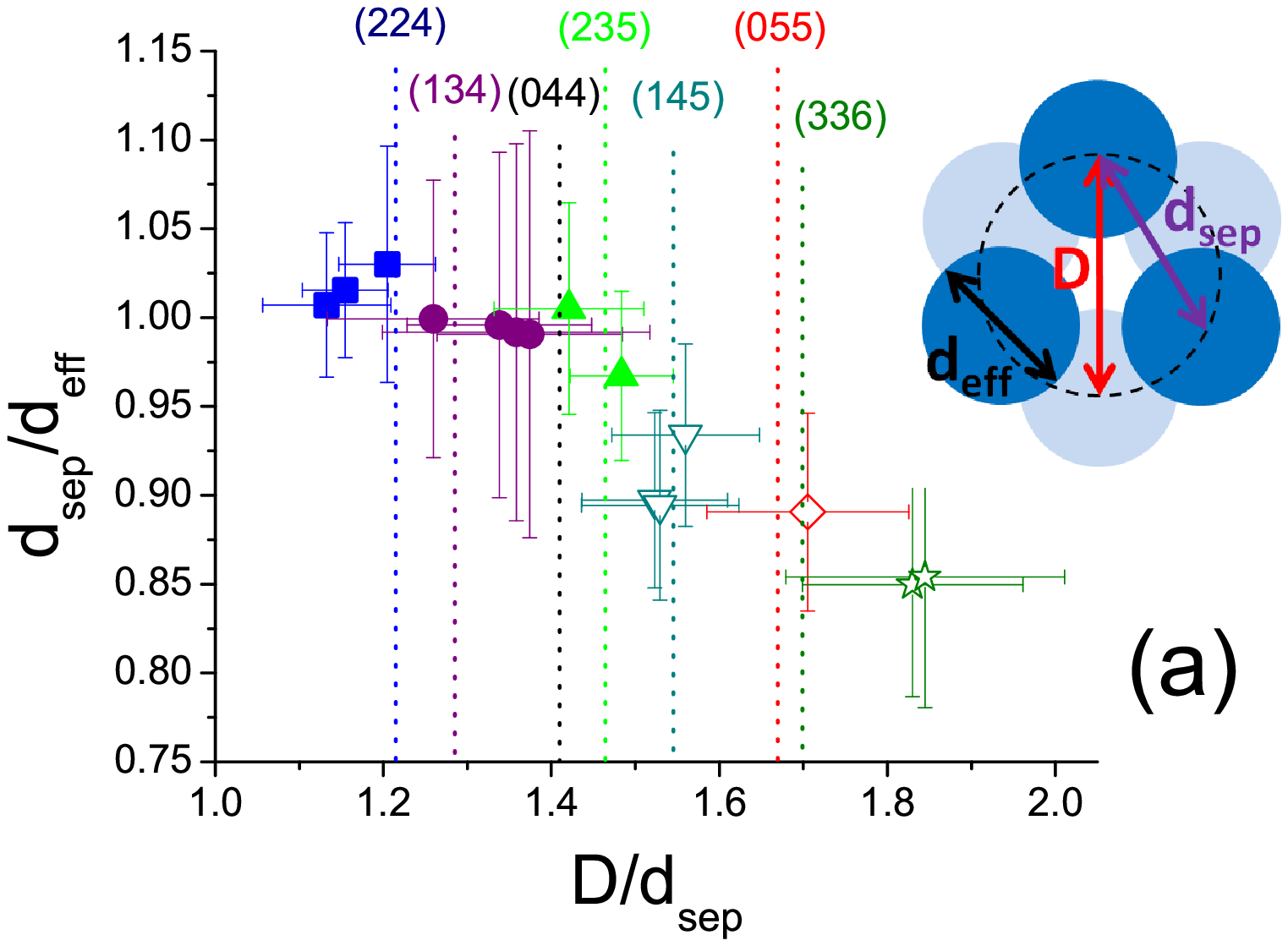}
\includegraphics*[viewport=20 615 563 755, scale=0.43]{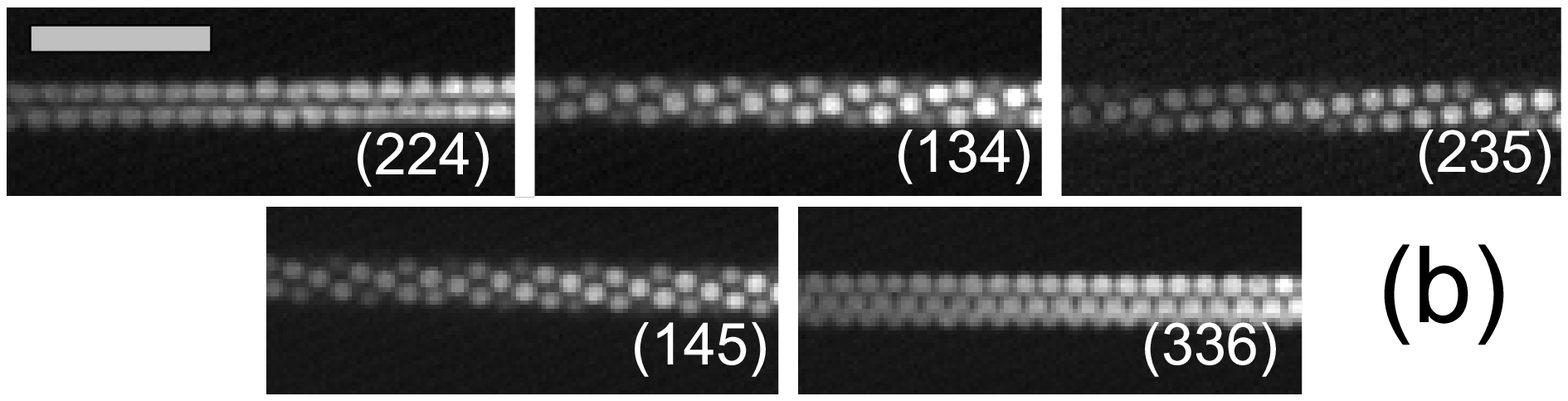}
\caption{(color online) (a) Geometrical characteristics of observed structures, (2,2,4) (blue $\square$), (1,3,4) (purple $\circ$), (2,3,5) (green $\bigtriangleup$), (1,4,5) (cyan $\bigtriangledown$), (0,5,5) (red $\Diamond$), (3,3,6) (dark green $\star$).  Filled symbols indicate structures with observable Brownian motion.  Dashed vertical lines indicate theoretical $D/d_{\hbox{\it sep}}$ values for predicted structures.  Inset: Cartoon of axial cross-section of a (336) structure with $d_{\hbox{\it sep}} > d_{\hbox{\it eff}}$.  (b) Confocal flourescence images of helical NIPA packings at high volume fraction.  Scale bar = 10$\mu$m.}
\end{figure}

At high densities, we observe crystalline helical packings with varying pitch and chirality dependent upon particle- and tube-diameter.  The observation of large ordered domains is consistent with the tendency for these nearly monodisperse particles to form uniform crystals in 2D [17] and 3D [20].  Particles in hard-sphere helical packings predicted for cylinders [5] have six nearest neighbors whose relative order along the tube axis corresponds to a characteristic set of three integers ($m, n, m+n$).  This notation for distinct helical crystalline structures is commonly used in phyllotaxy and is similar to the vector used to describe carbon nanotube chirality.  We verify such crystalline ordering from analysis of 3D confocal image stacks.  All varieties of predicted helical packings in the given range of tube-diameter-to-particle-diameter ratios were observed in 15 samples (Figure 1), with the exception of structure (0,4,4).  Additionally, the ratio $D/d_{\hbox{\it sep}}$, where $D$ is twice the average radial distance from the particle centers to the central axis of the tube and $d_{\hbox{\it sep}}$ is the average nearest neighbor particle separation, fell within experimental error of the predicted maximally packed hard-sphere values.

\begin{figure*}
\centering
\includegraphics*[viewport=0 0 595 410, scale=0.35]{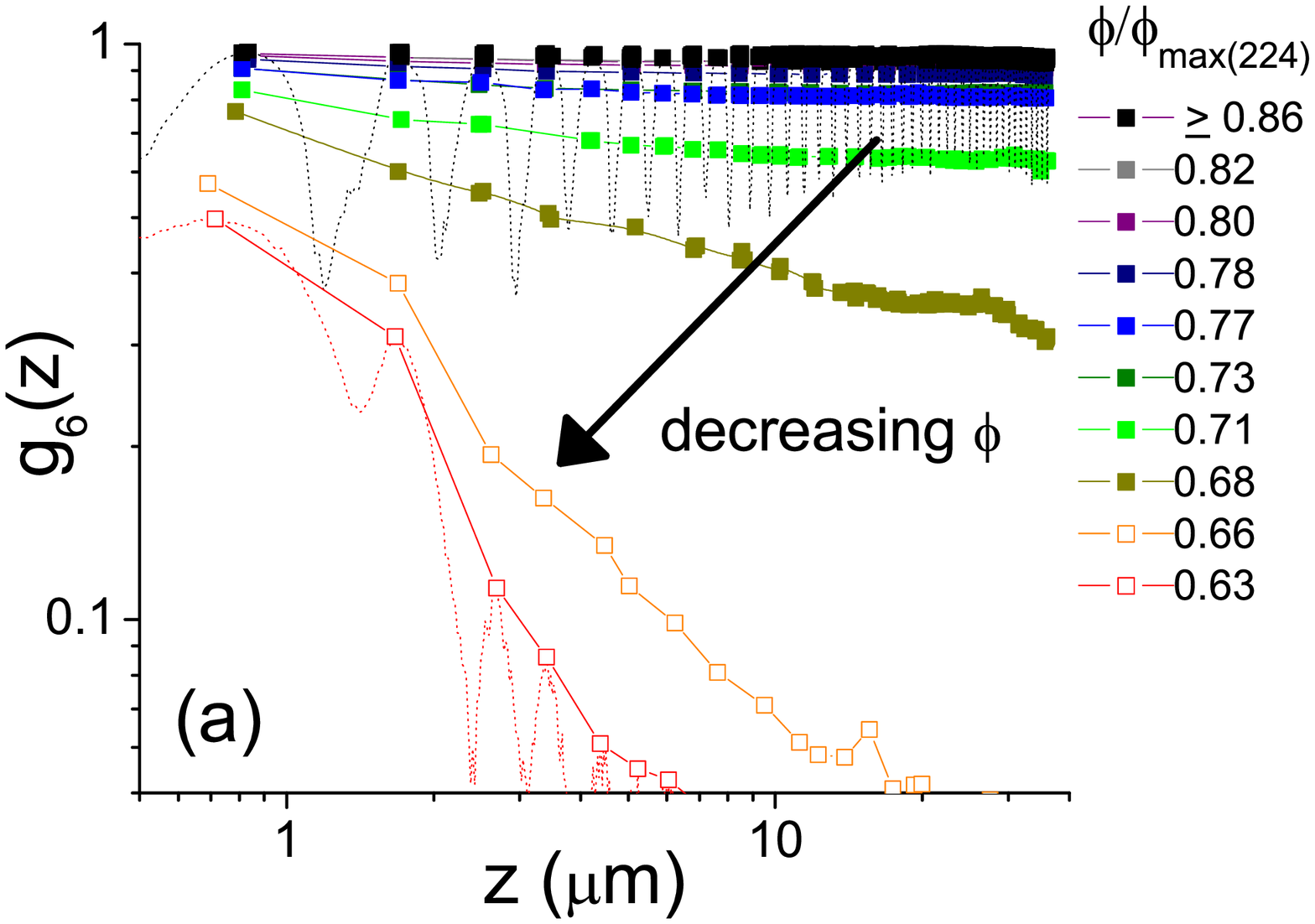}
\includegraphics*[viewport=0 0 595 450, scale=0.35]{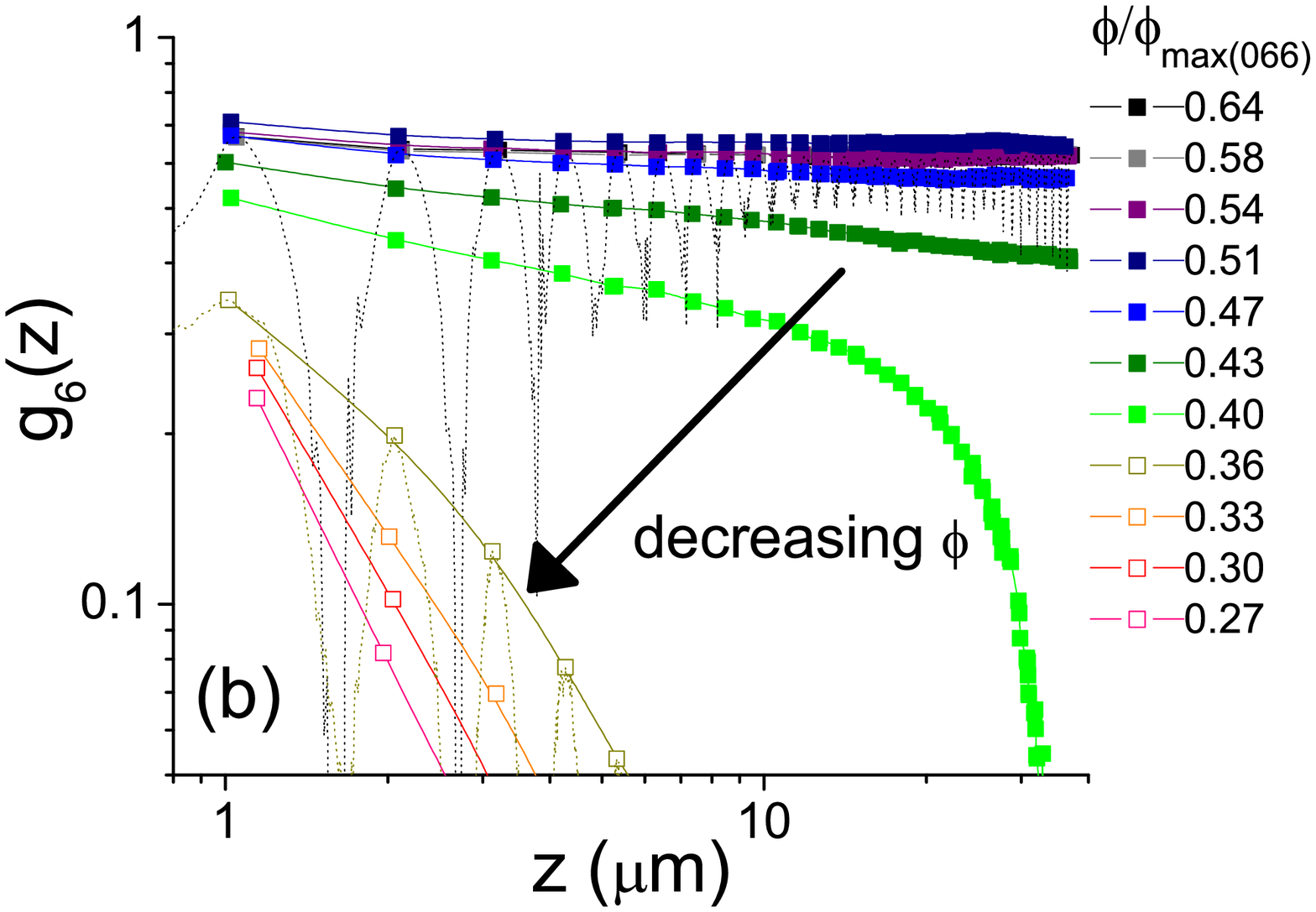}
\caption{(color online) Local maxima of orientational spatial correlation function $g_6(z)$ for samples with (2,2,4) packing (a) and (0,6,6) packing (b).  Dashed lines: Full correlation function at $\phi/\phi_{\hbox{\it max(224)}}$ = 0.86 and 0.63 (a) and $\phi/\phi_{\hbox{\it max(066)}}$ = 0.64 and 0.36 (b), where $\phi_{\hbox{\it max}}$ is the maximum volume fraction for a given packing of hard spheres with the observed particle spacing.  Empty symbols indicate samples with short-range order.  Oscillations arise from the periodic distribution of particle positions along the tube axis.  Full correlation functions were calculated for all volume fractions, but omitted for the sake of clarity.}
\end{figure*}

Ordered structures were found to exist over a range of volume fractions below maximal packing.  When the effective particle diameter $d_{\hbox{\it eff}}$ [23] was such that $d_{\hbox{\it sep}}/d_{\hbox{\it eff}} < 0.95$, the particles did not appear to move (i.e., motions greater than 0.2 $\mu$m were not observed during the 10-second scan).  In such cases, we consider the particles to be packed closer than their effective diameters.  For $d_{\hbox{\it sep}}/d_{\hbox{\it eff}} > 0.95 $, particles fluctuate significantly about their equilibrium positions; thus, thermal helical crystalline structures exist at volume fractions below close packing.

The volume fractions of such samples were then lowered further to determine if, when, and how the packings disorder to isotropic states.  Two uniformly packed samples were chosen and are presented here for careful analysis: a (2,2,4) packing of microspheres with diameter 1.71$\mu$m at 22$^\circ$C and a (0,6,6) packing of microspheres with diameter 1.23$\mu$m at 22$^\circ$C.  The sample temperature was increased in steps of 0.2-0.7$^\circ$C.  At each temperature step, after allotting ample time for the sample to reach thermal equilibrium (at least 5 minutes), videos of two-dimensional cross-sections of the packings were taken at 15 - 30 frames per second for approximately 5 minutes.  Though these two-dimensional videos lose some of the structural information available in three dimensional scans, they provide data at much higher speeds and yield better axial position tracking of particles in view.

A local orientational order parameter, $\psi_{6j} = \sum_k^{N_{nn}} e^{6i\theta_{jk}} / N_{nn}$, quantifies helical order in these systems.  Here, $\theta_{jk}$ is the angle between the axis of the tube and the bond between particles $j$ and $k$, and $N_{nn}$ is the number of nearest neighbors of particle $j$.  Though this order parameter is typically used for two-dimensional planar systems, it is acceptable to use in the analysis of two-dimensional slices of helical packings.  Helical packings are effectively two-dimensional triangular lattices wrapped into cylinders, and the observed cross-sections of these particular packings exhibit only slight variation from the ideal two-dimensional hexagonal lattice.

We examined the spatial extent of orientational order along the tube by calculating the orientational spatial correlation function, $g_6(z = |z_j-z_k|) = \langle \psi_{6j}^\ast \psi_{6k} \rangle$, where $z_j$ is the axial position of particle $j$.  As depicted in Figure 2, the resulting correlation functions decrease quickly at low volume fractions, as expected in a disordered state.  However, at high volume fractions, these functions exhibit long-range order within the experiment's field of view.  These experimental observations are consistent with an expectation of long-range orientational order.  Though long-range \textit{translational} order is impossible in one dimension at finite temperature [1], long-range \textit{orientational} order is possible, just as in the much-storied theory of two-dimensional melting [24]. 
\newcommand{\sinc}{{\rm sinc}}

One can show theoretically that long-range orientational order
persists even in this quasi-one-dimensional system by evaluating the
orientational correlation function $g_6$(r) using the isotropic
elasticity free energy [24,25] in an ``unwrapping'' of the particles
on the cylinder surface onto a two-dimensional infinitely long strip [26].  We assume
that particles are packed densely enough so that fluctuations along 
the radial direction of the cylinder may be neglected.  As 
$r\rightarrow\infty$, $g_6$ approaches a constant.  Thus, finite 
correlations exist at infinite distance, a hallmark of a phase with 
long-range order. Orientational order arises here through the
crystalline axes defined by unwrapping the cylinder, rather
than through an explicit additional mode, as in such systems studied in [27]. 
% [x] J.L. Lebowitz, J.K. Percus, and J. Talbot, J. Stat. Phys. 
% {\bf 49}, 1221-1234 (1987).

To our knowledge, the existence of long-range orientational order has not been characterized in previous studies of packing in cylindrical systems [11-15].  At low volume fractions, one expects long-range orientational correlations to disappear as observed in the experiment. However suggestive this combination of theory and experiment may be, we emphasize that further work is required to elucidate whether a 
generalization of KTHNY theory [24] is appropriate for this system.

To quantify the crossover from long-range to short-range order in the
experimental system, an average orientational order parameter $|\psi_6|$
is defined for each frame, where $\psi_6 = \frac{1}{N}\sum_j^N \psi_{6j}$.  The height of the first peak of the one-dimensional axial structure factor  $S(k_{max})$, where $S(k) = [N(N-1)]^{-1}\left|\sum\sum e^{ik|z_l - z_m|}\right|$, is used as a translational order parameter.  Here, $z$ denotes the axial position of each particle, $N$ is the number of particles in the field of view at a given time, and $k_{max}$ is chosen iteratively for each volume fraction.

\begin{figure}
\centering
\includegraphics*[viewport=35 12 553 400, scale=0.45]{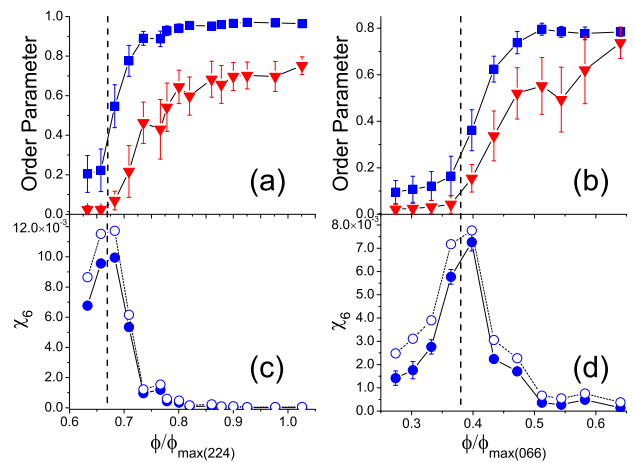}
\caption{(color online) Top: Mean values of translational (red $\blacktriangledown$) and orientational (blue $\blacksquare$) order parameters for (a) (2,2,4) and (b) (0,6,6) samples.  Bottom: Orientational susceptibilities for (c) (2,2,4) and (d) (0,6,6) systems, for sample size = 75 $\mu$m ($\circ$) and = $\infty$ ($\bullet$).  Dashed vertical lines at $\phi/\phi_{\hbox{\it max(224)}} = 0.67$ and $\phi/\phi_{\hbox{\it max(066)}} = 0.38$.}
\end{figure}

In Fig. 3(a,b), it is evident that both the average translational and orientational order parameters cross from an ordered state at high volume fraction to a disordered state at low volume fraction, though the change in $|\psi_6|$ is significantly sharper than the change in $S(k_{max})$.  The orientational crossover complements a recent simulation which finds a similiar crossover in hard sphere packings with decreasing density [14].

To characterize the fluctuations in orientational order, we calculate the orientational susceptibility $\chi_6 = \langle |\psi_6|^2 \rangle - \langle |\psi_6| \rangle^2$, where $\langle \rangle$ represents the time average.  Statistical effects of the finite size of the system are accounted for by calculating the susceptibility of different sub-segments of length $L$ in the system and extrapolating to the limit $L\rightarrow\infty$, similiar to the calculations in [17].  Plots of $\chi_6$ in Fig. 3(c,d) clearly demonstrate a peak in the orientational susceptibility.  The location of this peak coincides with the onset of both orientational and translational order in the system.  Translational susceptibilities were also calculated, but did not exhibit clear peaks or trends with respect to the order parameters or volume fraction.  We do not expect any transition-like behavior from translational susceptibilities due to arguments from [1] against long-range translational order, which apply in quasi-1D.

The existence of a diverging peak in the susceptibility of an order parameter typically indicates a phase transition [28].  However, it is difficult from the given data points to determine if this is truly a diverging peak, and whether it would indicate a first-order (asymmetrically diverging) or second-order (symmetrically diverging) phase transition.  Upon closer examination of the (0,6,6) sample, we observe coexistence of ordered and disordered domains for $\phi/\phi_{\hbox{\it max(066)}}$ = 0.47-0.40 (see Figure 4(a)).  We also observe the appearance of a small domain with dubious order in the (2,2,4) sample at $\phi/\phi_{\hbox{\it max(224)}}$ = 0.68.  The appearance of these coexisting domains is consistent with the spatial correlation functions exhibiting neither long-range nor short-range behavior at intermediate volume fractions in Figure 2 ($\phi/\phi_{\hbox{\it max(066)}}$=0.40, $\phi/\phi_{\hbox{\it max(224)}}$ = 0.68).  The presence of such solid-liquid coexistence has also been seen in recent simulations of hard spheres in cylinders [15].

In other samples, domains with different helical order often appear as volume fraction decreased (Figure 4b).  The appearance of these domains was difficult to quantify, since domains would grow, shrink and/or disappear with decreasing volume fraction.  The structures observed in coexisting states were those with most similar predicted linear densities and $D/d_{\hbox{\it sep}}$ values, consistent with recent hard sphere simulations [11,12].  This coexistence of ordered structures should not be confused with dislocation-mediated structural transitions theoretically studied in [6] and observed in [8] in athermal helical crystals, especially because of the difference in the observed sequence of coexisting structures.

\begin{figure}
\includegraphics*[viewport=25 355 200 455, scale=1.45]{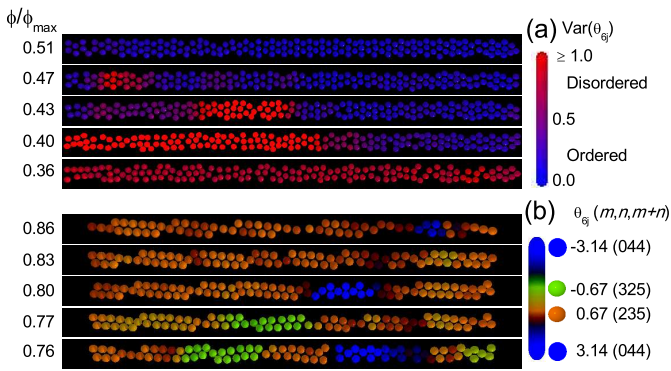}
\caption{(color online) Snapshots of particle tracks from systems exhibiting emergence of coexisting domains.  (a) (0,6,6) structure with emerging disordered domains.  Color represents local variance of the phase of $\psi_6$, $\langle \theta_{6j}^2 \rangle - \langle \theta_{6j} \rangle^2$, in 1.2 $\mu$m segments.  (b) (2,3,5) structure with emerging domains of (3,2,5) and (0,4,4) structures.  Color represents phase of $\psi_{6j}$, $\theta_{6j}$, which characterizes packing orientation.  Structures corresponding to each $\theta_{6j}$ given on right.}
\end{figure}

In summary, we created ordered helical packings of thermoresponsive colloids and observed the presence of long-range order resilient to thermal fluctuations.  Sharp crossovers from orientationally ordered to disordered phases with decreasing volume fraction were observed.  In addition, we find basic evidence for abrupt volume-fraction driven structure-to-structure transitions.  These findings raise and elucidate fundamental questions on the subject of melting in 1D.

\begin{acknowledgments}
We especially thank Yilong Han for clarifying discussions about experimental analyses, and we also thank Tom Haxton, Yair Shokef and Peter Yunker for enlightening discussions.  This work was supported by MRSEC grant DMR-0520020, NSF grant DMR-080488, and NASA grant NAG-2939.
\end{acknowledgments}

\noindent [1] L. Van Hove, Physica \textbf{16}, 137 (1950).\newline
[2] L. E. Hough \textit{et al.}, Science \textbf{325}, 456 (2009).\newline
[3] Y. Q. Zhou, C. K. Hall, and M. Karplus, Phys. Rev. Lett. \textbf{77}, 2822 (1996).\newline
[4] J. F. Douglas, Langmuir \textbf{25}, 8386 (2009).\newline
[5] R. O. Erickson, Science \textbf{181}, 705 (1973).\newline
[6] W. F. Harris and R. O. Erickson, J. Theor. Biol. \textbf{83}, 215 (1980).\newline
[7] N. Pittet, N. Rivier, and D. Weaire, Forma \textbf{10}, 65 (1995).\newline
[8] P. Boltenhagen and N. Pittet, Europhys. Lett. \textbf{41}, 571 (1998); N. Pittet \textit{et al.}, Europhys. Lett. \textbf{35}, 547 (1996); P. Boltenhagen, N. Pittet, and N. Rivier, Europhys. Lett. \textbf{43}, 690 (1998).\newline
[9] J. H. Moon \textit{et al.}, Langmuir \textbf{20}, 2033 (2004); F. Li \textit{et al.}, J. Am. Chem. Soc. \textbf{127}, 3268 (2005); M. Tymczenko \textit{et al.}, Adv. Mater. \textbf{20}, 2315 (2008).\newline
[10] A. N. Khlobystov \textit{et al.}, Phys. Rev. Lett. \textbf{92}, 245507 (2004); T. Yamazaki \textit{et al.}, Nanotechnology \textbf{19}, 045702 (2008).\newline
[11] G. T. Pickett, M. Gross, and H. Okuyama, Phys. Rev. Lett. \textbf{85}, 3652 (2000).\newline
[12] K. Koga and H. Tanaka, J. Chem. Phys. \textbf{124}, 131103 (2006).\newline
[13] M. C. Gordillo, B. Martinez-Haya, and J. M. Romero-Enrique, J. Chem. Phys. \textbf{125}, 144702 (2006).\newline
[14] F. J. Duran-Olivencia and M. C. Gordillo, Phys. Rev. E \textbf{79}, 061111 (2009).\newline
[15] H. C. Huang, S. K. Kwak, and J. K. Singh, J. Chem. Phys. \textbf{130}, 164511 (2009).\newline
[16] C. A. Murray and D. H. VanWinkle, Phys. Rev. Lett. \textbf{58}, 1200 (1987); K. Zahn, R. Lenke and G. Maret, Phys. Rev. Lett. \textbf{82}, 2721 (1999).\newline
[17] Y. Han \textit{et al.}, Phys. Rev. E \textbf{77}, 041406 (2008).\newline
[18] K. J. Strandburg, Rev. Mod. Phys. \textbf{60}, 161 (1988).\newline
[19] B. R. Saunders and B. Vincent, Adv. Colloid Interface Sci. \textbf{80}, 1 (1999); R. Pelton, Adv. Colloid Interface Sci. \textbf{85}, 1 (2000); L. A. Lyon \textit{et al.}, J. Phys. Chem. B \textbf{108}, 19099 (2004).\newline
[20] A. M. Alsayed \textit{et al.}, Science \textbf{309}, 1207 (2005).\newline
[21] H. Senff and W. Richtering, J. Chem. Phys. \textbf{111}, 1705 (1999); J. Wu, B. Zhou, and Z. Hu, Phys. Rev. Lett \textbf{90}, 048304 (2003); Y. Han \textit{et al.}, Nature (London) \textbf{456}, 898 (2008); Z. Zhang \textit{et al.}, Nature (London) \textbf{459}, 230 (2009); P. Yunker \textit{et al.}, Phys. Rev. Lett. \textbf{103}, 115701 (2009); J. Brijitta \textit{et al.}, J. Chem. Phys. \textbf{131}, 074904 (2009).\newline
[22] J. C. Crocker and D. G. Grier, J. Colloid Interface Sci. \textbf{179}, 298 (1996).\newline
[23] See Supplemental Material for details of particle diameter characterization.\newline
[24] D. R. Nelson and B. I. Halperin, Phys. Rev. B \textbf{19}, 2457 (1979).\newline
[25] N. D. Mermin, Phys. Rev. \textbf{176}, 250 (1968).\newline
[26] See Supplemental Material for additional details of this calculation.\newline
[27] Y. Kantor and M. Kardar, Phys. Rev. E \textbf{79}, 041109 (2009). \newline
[28] K. Binder, Rep. Prog. Phys. \textbf{50}, 783 (1987).\newline

\onecolumngrid
\pagebreak

\section{Helical Packings and Phase Transformations of Soft Spheres in Cylinders:
Supplemental Material}

\section{NIPA Microsphere Diameter Characterization}

Dilute suspensions of the same NIPA particles used in the tube packings were placed between glass coverslips such that the gap between coverslips was approximately the diameter of the NIPA particles, creating a quasi-2d monolayer.  Videos of these particles diffusing in two dimensions were taken using a 100$\times$ oil-immersion objective (N.A. = 1.4) at five temperatures from 24 to 28 $^\circ$C.  Particle centers were tracked using standard particle tracking routines [1].  The two dimensional pair correlation function $g(r)$ of the particle locations was then calculated from the particle tracks.  At each temperature, the approximate diameter of the particles taken to be the first value where $g(r) = 1/e$, since, in the first approximation, $g(r) ~ e^{-U(r)/k_BT}$, and the effictive diameter of particle is often taken as the value of $r$ for which $U(r) = 1 k_BT$.

Since previous studies have observed a linear relationship between diameter and temperature for these particles in this temperature range [2], we then take a linear fit of these data points to find a functional relationship between effective particle diameter and temperature.  For the larger species used in this experiment, we find the relationship $d_{eff} = 2.52\mu$m$ - 0.037(\mu$m$/^\circ C) \times T$, and for the smaller species, we find $d_{eff} = 2.41\mu$m$ - 0.054(\mu$m$/^\circ C) \times T$.

For videos of two dimensional cross sections of particle packings in cylinders, the nearest neighbor particle spacing $d_{sep}$ was calculated from axial particle spacings and theoretical helical packing values.  The volume fraction ratios of the tube packings were then calculated as $\phi/\phi_{max} = (d_{eff}/d_{sep})^3$.

\begin{figure*}[h]
\centering
\includegraphics*[viewport=25 0 520 405, scale=0.35]{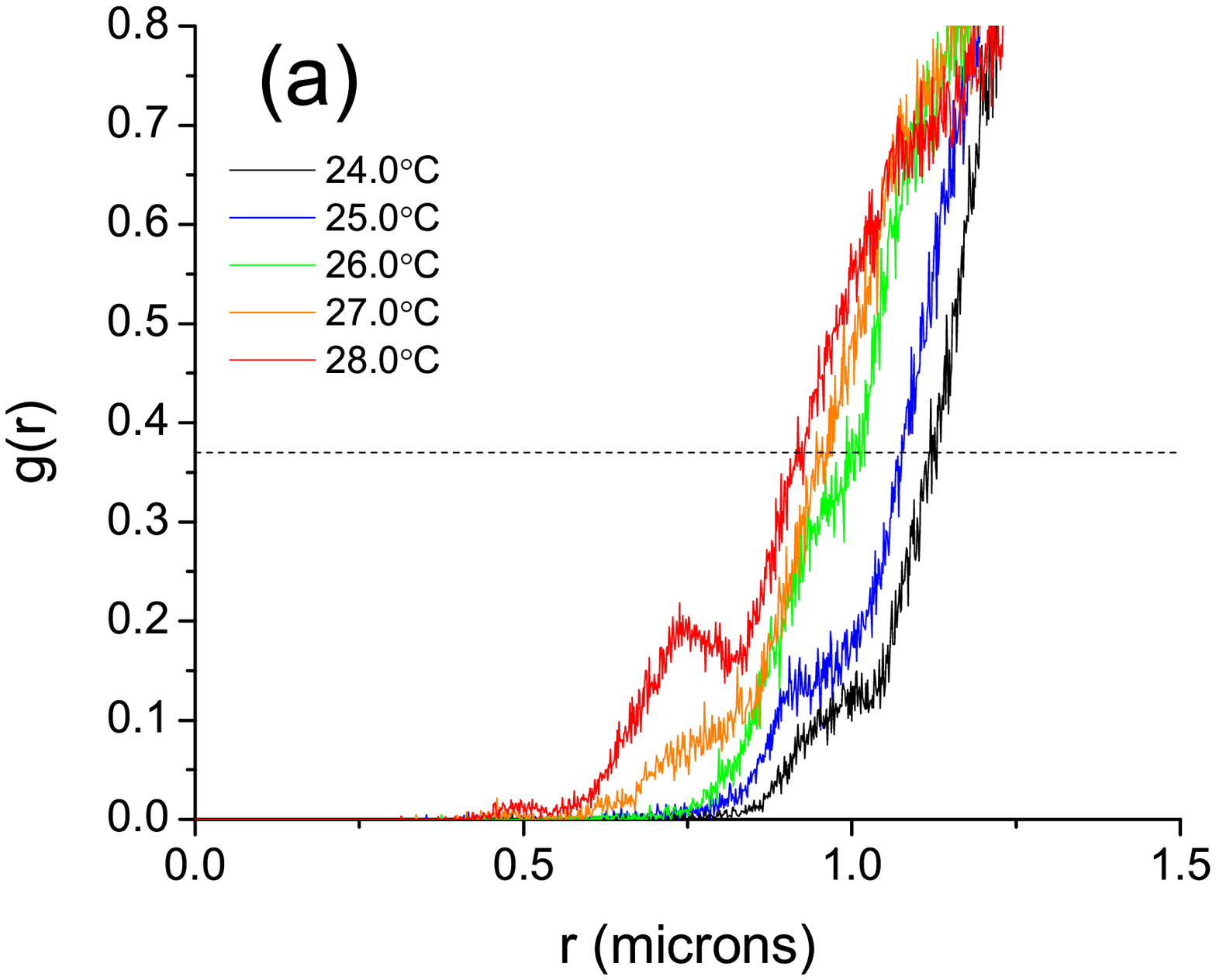}
\includegraphics*[viewport=25 0 520 405, scale=0.35]{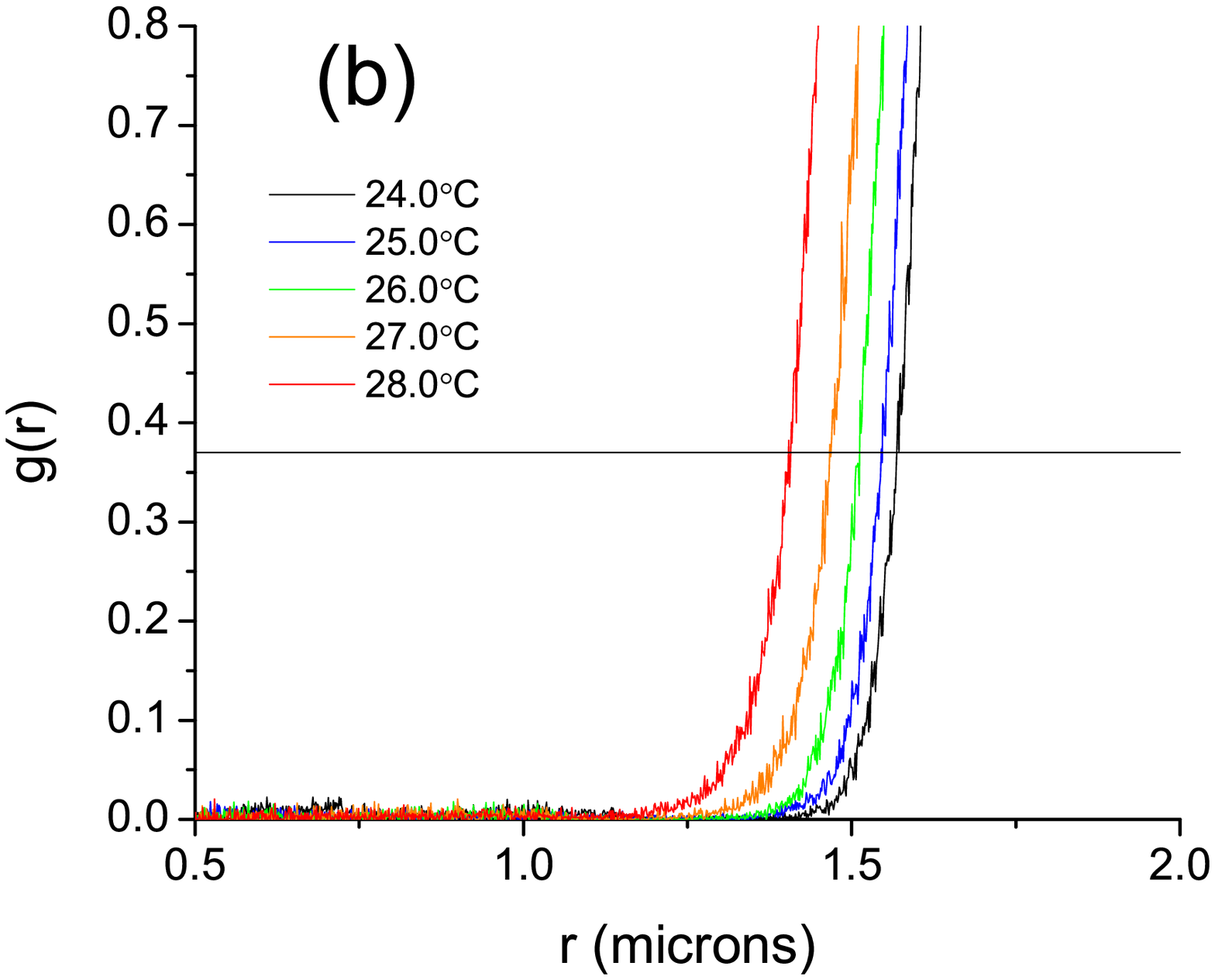}
\caption{(a) Two-dimensional correlation functions for (a) smaller and (b) larger NIPA particles at different temperatures.  Horizontal dashed lines at $g(r) = 1/e$.}
\end{figure*}

\begin{figure}
\centering
\includegraphics*[viewport=25 0 520 405, scale=0.35]{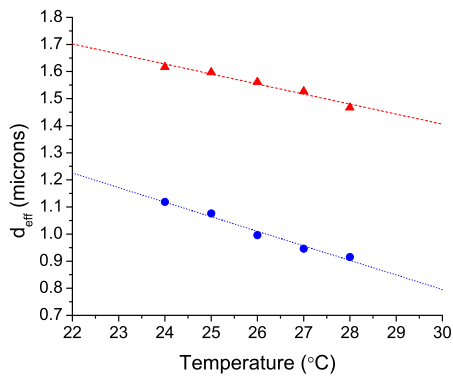}
\caption{Approximate diameters for the smaller (blue $\bullet$) and larger (red $\blacktriangle$) NIPA particles at different temperatures, with linear fits.}
\end{figure}

\section{Identification of Structures with Predicted Helical Packings}

We describe the predicted helical packings with a commonly used three-index notation, ($m, n, m+n$) [3].  If we consider any single particle in a such an ideal close-packed system of hard spheres, we notice that each particle six nearest neighbors.  We can thus think about such a packing as a two-dimensional triangular lattice wrapped around a cylinder.  The indices then indicate the unit vectors in the unwrapped triangular lattice connecting any point to itself in the helical structure.  Alternatively, one notices that the three indices indicate the relative distances of a particle's nearest neighbors along the axis of the cylinder.  For example, if any given particle's nearest neighbors are the second, third or fifth closest particles in the axial direction, it would exist in a (2,3,5) packing. 

After identifying particle centers in 3D using common particle tracking routines [1], the nearest neighbors of each particle in an image are identified as those closer than the far end of the first maximum in the 3D pair correlation function $g(r)$ (see Figure 3b).  After an individual particle in the packing is selected, all other particle are given integer values based on the order of their axial distance from the selected particle in each direction.  The integer values of the nearest neighbors are then recorded.  This process is repeated for every particle in the structure.  This creates histograms of the relative axial order of the neighboring particles.  The peaks in these histograms then identify the integers ($m, n, m+n$) describing the ideal packing (see Figure 3c).

\begin{figure}[h]
\centering
\includegraphics*[viewport=100 0 375 405, scale=0.30]{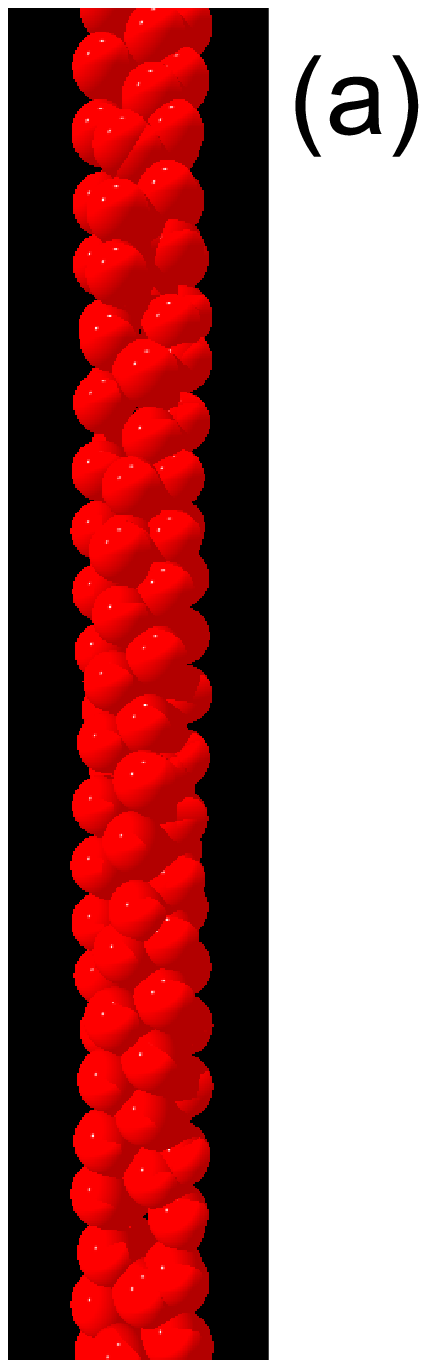}
\includegraphics*[viewport=25 0 520 405, scale=0.30]{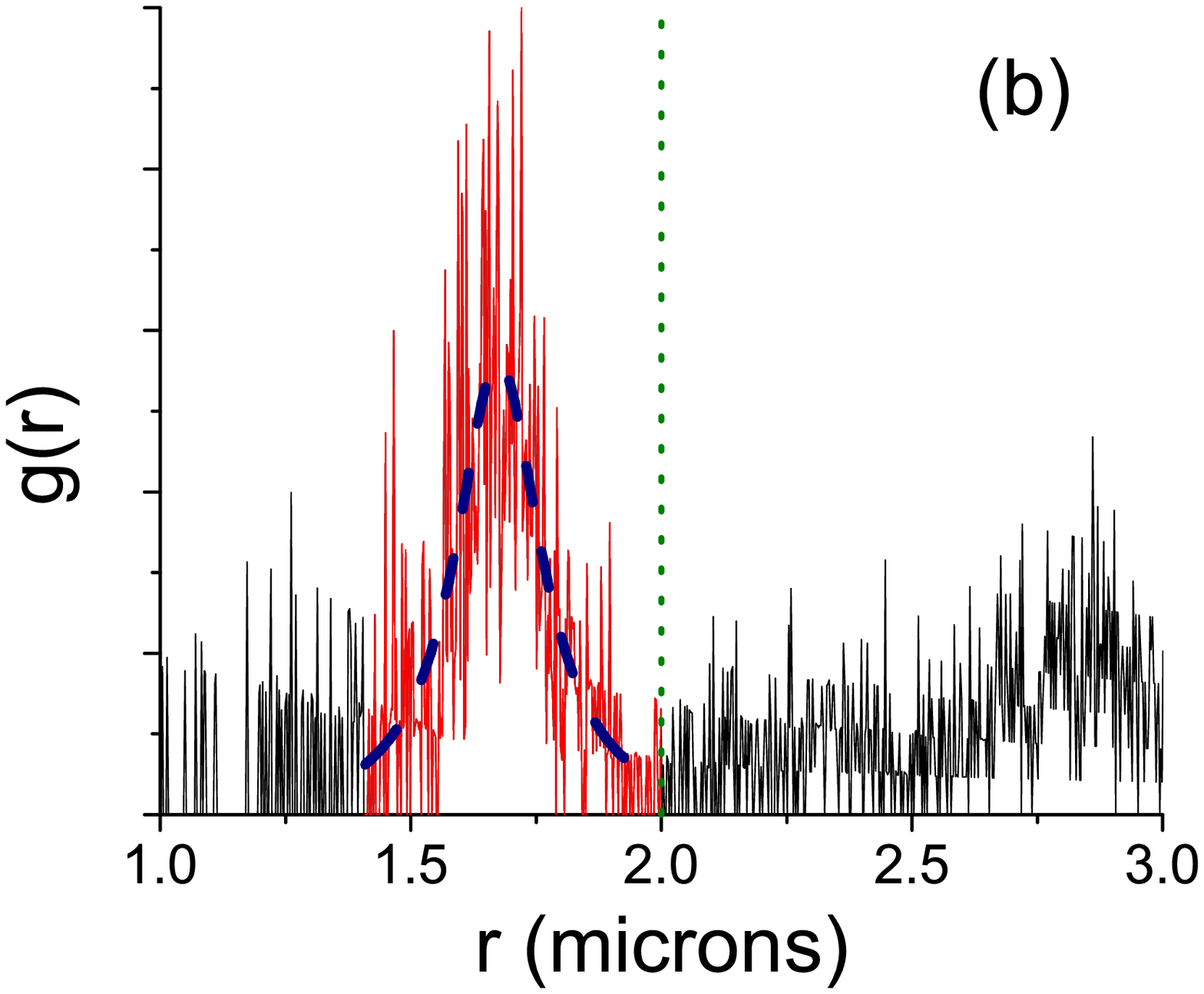}
\includegraphics*[viewport=25 10 310 285, scale=0.50]{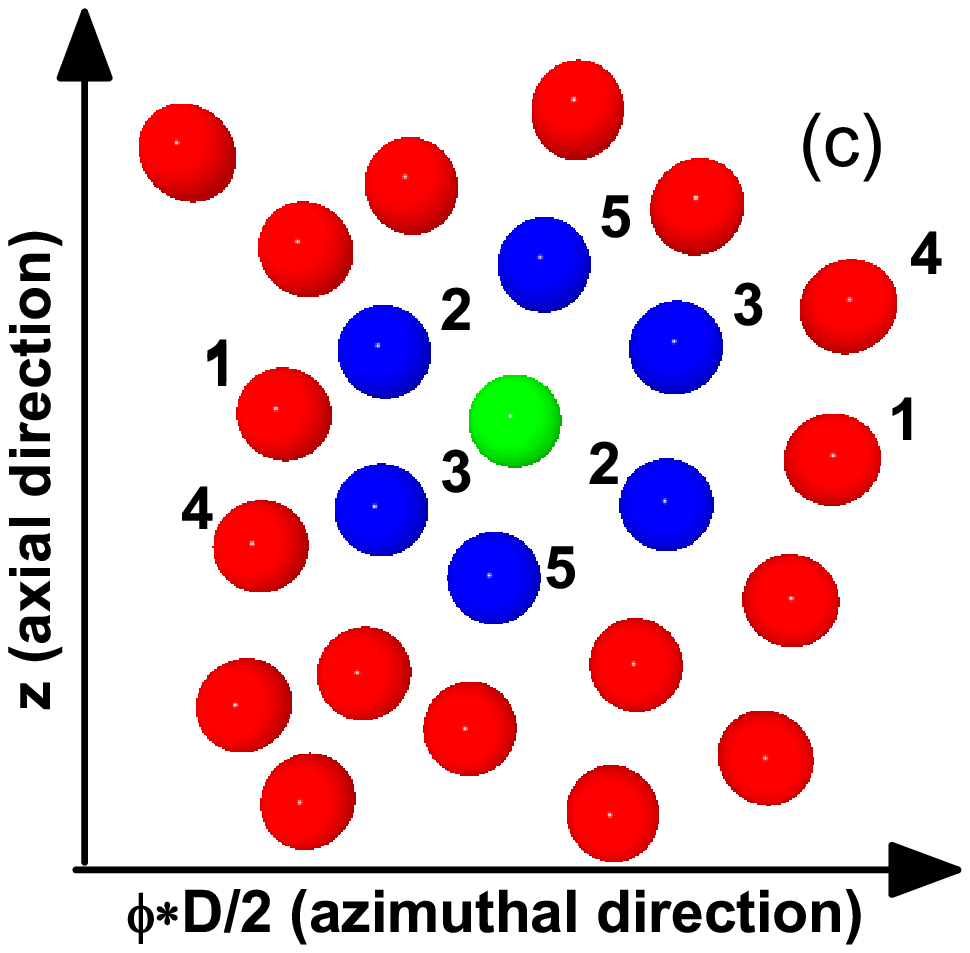}
\includegraphics*[viewport=20 0 520 405, scale=0.30]{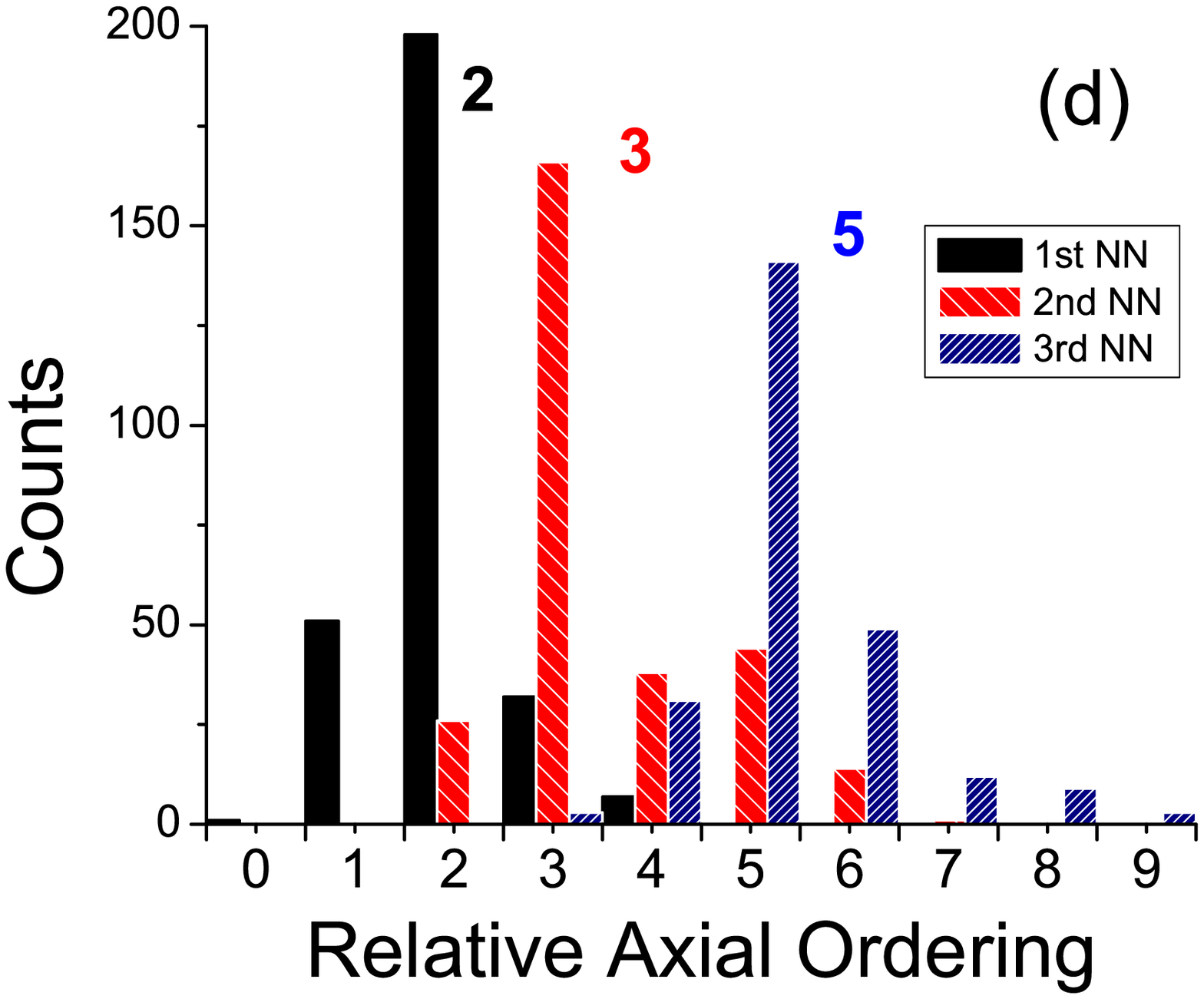}
\caption{Steps involved in categorizing structures and quantifying structural parameters of a sample (2,3,5) structure.  (a) Three-dimensional reconstruction of microsphere packing from tracking of confocal image stack.  (b) 3D pair correlation function from particle positions.  First peak in $g(r)$ highlighted in red, with Lorentzian peak fit in blue dashed line.  Vertical dotted line indicates nearest neighbor cutoff.  (c) Axial and azimuthal positions of tracked particles in a short axial section.  Selected particle position highlighted in green, with nearest neighbors highlighted in blue.  Relative axial distance from the selected particle listed to side.  Note that the nearest neighbors have relative axial order of 2, 3 and 5.  (d) Histogram of relative axial order for 1st, 2nd and 3rd closest nearest neighbors in the axial direction.  Peaks at values of 2, 3 and 5 indicate that this is a (2,3,5) packing.}
\end{figure}

\section{Estimation of Structural Parameters from 3D Tracks}

The average inter-particle spacing $d_{sep}$ in a 3D track is determined from the 3D pair correlation function.  Since each structure has on the order of 100 particles, the resolution of said correlation function is fairly poor.  Thus, we determine $d_{sep}$ by fitting a Lorentz peak, $f(r) = f_0 + A(w/(4(r-r_0)^2+w^2))$ to the first peak in $g(r)$ (see Figure 3b).  The center of this function, $r_0$, gives us the value for $d_{sep}$.  The value for structure diameter $D$ was taken as twice the average value of the radial positions of the particles in the structure. 

\section{Elastic approach to long range orientational order in a cylindrical system of colloids}

%Consider a system of colloids packed into a cylinder.  In experiment
%it was observed that under certain conditions, by varying the volume
%fraction of the spheres, one could observe a transition from a state
%where the spheres appeared to lie in an ordered packing to a totally
%disordered state.  Additionally, a peak in the orientational
%susceptibility was measured.  What ``phases'' exist in this system?
%What kind of transition is this?  A true phase transition, or merely
%the low-dimensional / finite-size shadow of one?  

In one- and two-dimensional systems of particles, there can be no long 
range positional order, however, there can be long range orientational
order in two dimensions [4].  This fact is key to the
KTHNY theory [5-7] of continuous melting in two dimensions.
%Qualitatively speaking, at low temperatures below $T_m$, the two
%dimensional solid has quasi-long range translational order and long
%range orientational order.  At $T_m$, dislocation pairs unbind and the
%system is a hexatic, with exponentially suppressed translational order
%and now quasi-long range orientational order.  At $T_i$, disclination
%pairs unbind and the system is isotropic, with no long range order of
%any type.

We point out here that the same arguments used by Nelson and Halperin
in [6] applied to a quasi-{\em one}-dimensional system, show
that there is long range orientational order in such systems at 
low temperatures.  The system we consider is a two-dimensional box which 
is infinite in the $x$ direction and periodic with length $L$ in the 
$y$ direction.  Note that such a system system applies to the problem of spherical particles in a
cylinder if we assume that at high densities, the outer layer of spheres in the cylinder behave as if they were disks lying in a two-dimensional strip with periodic boundary conditions -- that is, 
we ``unwrap'' the spheres onto a plane and neglect the radial (which become out-of-plane) fluctuations.  This maps $\psi_6$ defined in the quasi-one-dimensional system approximately onto the $\psi_6$ that was measured on the surface of the cylinder. 

We'll calculate a few quantities in the
low-temperature (``solid'' phase) by using an isotropic Lam\'e 
elasticity free energy.  The following expression is appropriate for
the elasticity of a triangular lattice with the given geometry:
\begin{equation}
F=\frac{1}{2}\int_{-\infty}^\infty dx\int_{0}^{L} dy 
\left[\lambda u_{ii}^2 +2\mu u_{ij}u_{ij}\right]
\end{equation}

Here $u_{ij}=\frac{1}{2}\left(\partial_i u_j+\partial_j u_i\right)$ is
the displacement from the (zero-temperature, hexagonal) ordered state.  
This free energy assumes that there exist nonzero elastic moduli
$\lambda$ and $\mu$, a natural assumption we must make.  

%As in two dimensions, at finite temperatures, there can be no long range 
%positional order in this system, so fluctuations drive the
%correlations of $\psi_G=e^{iG\cdot u}$ (the density amplitude at
%reciprocal lattice vector $G$) to zero at large $x$.  This can
%be seen by evaluating $\langle \psi_T(x)\psi_T(0)\rangle$:
%\begin{align*}
%\langle \psi_G(x)\psi_G^*(0)\rangle&=\langle e^{iG\cdot (u(x)-u(0))}
%\rangle\\
%&=\exp\left(-G^2[\langle u(0)^2\rangle-\langle
%u(x)u(0)\rangle]\right)\\
%&=\exp\left(-G^2\frac{k_BT}{L}\sum_{n=-N}^N\int_{-\Lambda}^\Lambda 
%\frac{dq_x}{2\pi} \right)
%\end{align*}

To calculate the correlation function of orientational order, we first
define $\psi_6=e^{6i\theta}$, where $\theta$ measures the bond angles
between nearest neighbor particles (the factor of 6 arises from the
triangular symmetry of the lattice). Furthermore, we will use the fact
that the angle $\theta$ can be shown to be $\frac{1}{2}\nabla\times u$.  

The fact that the theory is quadratic allows us to take two shortcuts.
First, we can evaluate $\langle \psi_6(r)\psi_6(0)\rangle$ (where
$r=(x,y)$) as the exponential of an average. Second, this average can
be evaluated simply since the inverse of the elasticity dynamical
matrix is known [6,8] to be $D_{jn}^{-1}=\left\{\frac{1}{\mu q^2}
\left(\delta_{jn}-\frac{q_jq_n}{q^2}\right)
+\frac{1}{\lambda+2\mu}\frac{q_jq_n}{q^4}\right\}$. 

It may be worthwhile to point out that we can guess
the result from ``counting powers of $q$''.  We can see that
translational order is destroyed in this system by estimating the
scaling of the (Fourier-space) correlation function of $u$ at low $q$.
To do this, we integrate $D^{-1}_{jn}$ in one dimension.  But
$D^{-1}_{jn}$ scales as $q^{-2}$, so the result $\int q^{-2}dq$ scales
like $1/q$.  This diverges at low $q$, meaning that 
long-wavelength fluctuations drive positional correlations to zero.  

However, since $\theta$ is related to the {\em derivative} of
$u$, the (Fourier-space) correlation function of $\theta$ will involve 
an extra factor of $q$ for each $\theta=\nabla\times u$.  We thus expect 
the two-point correlation function to scale as 
$\int dq(q^2)(q^{-2})\sim q$, which is finite at low $q$,
indicating long range order at infinity.

We now proceed with a more detailed calculation of the real-space
orientational order correlation function:
\begin{align}
\nonumber \langle \psi_6(r)\psi_6(0)\rangle&=\exp[-36
(\langle\theta^2(0)\rangle-\langle\theta(r)\theta(0)\rangle)]\\ \nonumber
&=\exp\left[-\frac{9k_BT}{2\pi L}\epsilon_{ij}\epsilon_{mn}
\sum_{n=-N}^N\int_{-\Lambda}^\Lambda dq_x q_iq_m
\left(1-e^{i(q_x x+2\pi n y/L)}\right)D_{jn}^{-1} \right]\\ \nonumber
&=\exp\left[-\frac{9k_BT}{2\pi\mu L}
\sum_{n=-N}^N\int_{-\Lambda}^\Lambda dq_x
\left(1-e^{i(q_x x+2\pi n y/L)}\right) \right]\\ \nonumber
&=\exp\left[-\frac{9k_BT}{2\pi\mu L}
\sum_{n=-N}^N\left(2\Lambda-e^{i2\pi n y/L}\frac{e^{i\Lambda
x}-e^{-i\Lambda x}}{ix}\right) \right]\\
&=\exp\left[-\frac{9k_BT}{2\pi\mu L}
\left(2\Lambda(2N+1)-\frac{\sin(2\pi(N+1/2)y/L)}{\sin(\pi y/L)}
\frac{2\sin(\Lambda x)}{x}\right) \right]
\end{align}

We now let the cutoff $\Lambda$ be $2\pi(N+1/2)/L$ and use the sinc
function ($\text{sinc }t=(\sin t)/t$):

\begin{align}
\nonumber \langle \psi_6(r)\psi_6(0)\rangle
&=\exp\left[-\frac{9k_BT}{\pi^2\mu }
\left(\Lambda^2-\Lambda^2\frac{\pi y}{L}
\frac{\sin(\Lambda y)}{\sin(\pi y/L)\Lambda y}
\frac{\sin(\Lambda x)}{\Lambda x}\right) \right]\\
&=\exp\left[-\frac{9k_BT\Lambda^2}{\pi^2\mu }
\left(1-\frac{\text{sinc}(\Lambda x)
\text{sinc}(\Lambda y)}{\text{sinc}(\pi y/L)}
\right)\right]
\end{align}

For large $x$, this approaches the constant
$\exp\left(-\frac{9k_BT\Lambda^2}{\pi^2\mu }\right)$, and hence this
quasi-one dimensional system has long range orientational order.  This result explains
our experimentally observed long range correlations in the orientational order parameter at high densities.

We emphasize that (3) should not be fit to experimental or simulation data in the present form for the following reasons.  First, the elastic field theory describes the sphere systems at large wavelengths. This approach is fine for the purposes of searching for long-range orientational order, but is inappropriate for the extraction of quantitative correlation functions.

Another issue is that the precise functional form of (3) arises from the cutoff scheme chosen. Ideally, one would apply a physical theory which describes the higher wavelength modes and how the system responds to them (in the case of spheres, part of this scheme could be a density functional-like theory), which would be more suitable for capturing the behavior of the correlation function over short distances than the hard cutoff that we used.

Thus the primary experimentally accessible result for comparison to theory is the fact that the correlation function remains finite at infinity.

%If we just kept the $z$-direction in this
%problem, it would become impossible to analyze the orientational order
%of this system.

%\begin{figure}
%\centering
%\includegraphics[width=10cm]{rothdurianpolarsq}
%\caption{Shape profiles for equation \ref{eqn:ev2} on a square, at
%$t=0$ (blue) and $t=0.5$ (red)}
%\label{fig:rothdurianpolarsq}
%\end{figure}

%\paragraph*{Acknowledgments.}
%I'd like to thank Randy Kamien for bringing me aboard this project and 
%suggesting this approach, as well as working out some details with me.  
%Also many thanks to Matt Lohr for access to data and many useful 
%conversations.

\begin{bibliography}
\noindent[1] J. C. Crocker and D. G. Grier, J. Colloid Interface Sci. \textbf{179}, 298 (1996).\newline
[2] A. M. Alsayed \textit{et al.}, Science \textbf{309}, 1207 (2005); Y. Han \textit{et al.}, Phys. Rev. E \textbf{77}, 041406 (2008); Y. Han \textit{et al.}, Nature (London) \textbf{456}, 898 (2008); Z. Zhang \textit{et al.}, Nature (London) \textbf{459}, 230 (2009); P. Yunker \textit{et al.}, Phys. Rev. Lett. \textbf{103}, 115701 (2009).\newline
[3] R. O. Erickson, Science \textbf{181}, 705 (1973); W. F. Harris and R. O. Erickson, J. Theor. Biol. \textbf{83}, 215 (1980).\newline
[4] N. D. Mermin, Physical Review \textbf{168}, 250 (1968).\newline
[5] J.M. Kosterlitz and D.J. Thouless, J Phys C \textbf{6}, 1181 (1973).\newline
[6] D.R. Nelson and B.I. Halperin, Phys. Rev. B \textbf{19}, 2457 (1979).\newline
[7] A.P. Young, Phys. Rev. B \textbf{19}, 1855 (1979).\newline
[8] P. Chaikin and T. C. Lubensky, \textit{Principles of Condensed Matter Physics} (Cambridge University Press, Cambridge, England, 2006).\newline
\end{bibliography}
\end{document}